\documentclass[11pt, a4paper]{article}
\usepackage[height=8.85in,width=6.4in]{geometry}

\usepackage{amsmath}
\usepackage{mathtools}
\usepackage{amsfonts}
\usepackage{amssymb}
\usepackage{slashed}
\usepackage{braket}
\usepackage{bm}
\usepackage{comment}
\usepackage{cite}
\usepackage[utf8]{inputenc}
\usepackage[footnotesize]{caption}
\input{colordvi.tex}
\usepackage{feynmp}

\usepackage{fourier}

\usepackage[pdftex]{graphicx}
\usepackage{bm}
\usepackage[pdftex]{hyperref}
\usepackage[svgnames]{xcolor}
\definecolor{phthaloblue}{rgb}{0.0, 0.06, 0.54}
\hypersetup{
    colorlinks=true,
    linkcolor=phthaloblue,
    citecolor=blue,
    filecolor=blue,
    urlcolor=phthaloblue,
    }

\def\({\left(}
\def\){\right)}
\def\[{\left[}
\def\]{\right]}

\def\nn{\nonumber \\}

\def\Mpl{M_{\rm pl}}
\def\lmk{\left(}
\def\rmk{\right)}
\def\lkk{\left[}
\def\rkk{\right]}

\newcommand{\eq}[1]{Eq.~(\ref{#1})}
\newcommand{\bel}[1] {\begin{equation}\label{#1}}
\newcommand{\beal}[1] {\begin{eqnarray}\label{#1}}
\newcommand{\be}{\begin{equation}}
\newcommand{\ee}{\end{equation}}
\newcommand{\bea}{\begin{array}} 
\newcommand{\eea}{\end{array}}

\def\del{\partial}

\newcommand{\GeV}{\  {\rm GeV} }

\newcommand{\dd}{\mathrm{d}}

\begin{document}

\begin{titlepage}

\begin{center}

\hfill DESY 19-014

\vskip 1.2in

{\Huge \bfseries 
Dark Matter Production\\ during the Thermalization Era\\
}
\vskip .8in

{\Large Keisuke Harigaya$^\natural$, Kyohei Mukaida$^\sharp$, Masaki Yamada$^\flat$}

\vskip .3in
\begin{tabular}{ll}
$^\natural$&\!\!\!\! \emph{School of Natural Sciences, Institute for Advanced Study, Princeton, NJ 08540, USA}\\
$^\sharp$&\!\!\!\! \emph{DESY, Notkestra{\ss}e 85, D-22607 Hamburg, Germany}\\
$^\flat$&\!\!\!\! \emph{Institute of Cosmology, Department of Physics and Astronomy, Tufts University,}\\ 
&\!\!\!\!\emph{ Medford, MA  02155, USA}
\end{tabular}

\end{center}
\vskip .5in

\begin{abstract}
\noindent
We revisit the non-thermal dark matter (DM) production during the thermalization and reheating era after inflation. The decay of inflaton produces high-energy particles that are thermalized to complete the reheating of the Universe. Before the thermalization is completed, DM can be produced from a collision between the high-energy particles and/or the ambient plasma. We calculate the DM abundance produced from these processes for the case where the cross section of the DM production is proportional to the $n$-th power of the center of mass energy. We find that the collision between the high-energy particles is almost always dominant for $n \gtrsim 4$ while it is subdominant for $n \lesssim 2$. The production from the ambient plasma is dominant when $n \lesssim 3$ and the reheating temperature is of the order of or larger than the DM mass. The production from a collision between the high-energy particle and the ambient plasma is important for $n \lesssim 2$ and the reheating temperature is much lower than the DM mass. 
\end{abstract}

\end{titlepage}

\section{Introduction}

Weakly interacting dark matter (DM) with its abundance determined by the freeze-out process~\cite{Lee:1977ua} has been intensively studied, especially in connection with solutions to the hierarchy problem. 
The Large Hadron Collider so far has not found any particles beyond the Standard Model (SM), which encourages us to also consider broader classes of dark matter candidates and production mechanisms. In this paper, we focus on the non-thermal DM production during the thermalization of inflaton decay products, 
which has been pointed out in Ref.~\cite{Harigaya:2014waa}. We will not attempt to apply the production mechanism to a specific model, but the computation presented in this paper should be useful for many models.

In most models of inflation and dark matter, 
the mass of inflaton is many orders of magnitude larger than the electroweak scale. 
For example, 
the inflaton mass in the Starobinsky inflation model~\cite{Starobinsky:1980te} 
is of order $10^{13} \GeV$. 
The heavy inflaton decays into high-energy particles in the SM 
so that the Universe is reheated after the end of inflation. 
The decay products, whose energy is of the same order as 
the inflaton mass, lose their energy to be thermalized to form thermal plasma.

In the very early stage of reheating, the parametric resonance may take place~\cite{Traschen:1990sw,Kofman:1994rk,Shtanov:1994ce,Kofman:1997yn}. There high-energy particles are produced non-perturbatively owing to the high occupation number of the inflaton.
Particle production in this stage has been investigated in the literature~\cite{Felder:1998vq}.
However, typically, if the reheating proceeds via a very tiny coupling of the inflaton with radiation, \textit{e.g.}, Planck-suppressed operators, the resonance is soon shut off by the cosmic expansion.
Afterwards, the reheating process is well described perturbatively.
Even in this regime, there is a non-trivial way to create DM which we will investigate in this paper.

The thermalization process via a perturvative reheating has been studied in detail in Refs.~\cite{Harigaya:2013vwa, Mukaida:2015ria} (see also Refs.~\cite{Davidson:2000er, Jaikumar:2002iq} for earlier works and Refs.~\cite{Arnold:2002zm, Kurkela:2011ti, York:2014wja, Kurkela:2014tea,Kurkela:2014tla} in a different context). 
The high-energy particles emit soft particles, which have relatively small energy, via inelastic scattering processes. 
The soft particles interact among them to form thermalized ambient plasma.
As the temperature of the ambient plasma increases, 
the interaction between the high-energy particles and the soft particles 
becomes efficient. 
Eventually, the high-energy particles become completely thermalized. 
This is the time when the temperature is maximum after inflation. 
Still, 
the inflaton does not decay completely, 
so that the high-energy decay products are continuously produced from the inflaton decay. 
When the Hubble parameter becomes comparable to the inflaton decay rate, 
the reheating process is completed and the Universe becomes dominated by the thermal plasma.

The most important process in the thermalization era is 
the inelastic scattering process.
This process results in 
the splitting of a high-energy particle, whose energy is larger than the temperature of the ambient plasma, into two high-energy particles.
Here we assume $m_\phi > T$.\footnote{
If the temperature of ambient plasma becomes larger than the mass of inflaton, i.e. $m_\phi \ll T$, we cannot compute the inflaton decay rate by just computing it in zero temperature field theory.
This is because the dispersion relation of decay products get modified, and in particular, they acquire the thermal mass. For $m_\phi < T$, it is possible that the thermal mass exceeds the inflaton mass term, indicating that the inflaton cannot decay into quasi particles in thermal plasma.
However, it does not mean that inflaton cannot convert its energy into the plasma because inflaton can lose its energy by (multiple) scatterings from these quasi particles~\cite{Drewes:2010pf,Mukaida:2012qn,Mukaida:2012bz,Drewes:2013iaa}.
}
A high-energy particle with an energy as large as the inflaton mass $m_\phi$ 
splits into two particles whose typical energy is about the half of the initial one. 
Continuing this splitting process, 
the original high-energy particle 
splits into $N$ particles with a typical energy of order $m_\phi / N$. 
This process continues until the energy drops to the temperature of the ambient plasma. 
This means that numerous high-energy particles are produced during the thermalization process from the decay of a single inflaton. 
The DM may be efficiently produced from those high-energy particles 
before they are completely thermalized~\cite{Harigaya:2014waa}.

The thermalization via the splitting does not occur instantaneously.
There are two important consequences of the finiteness of thermalization time scale. 
Frist, 
the maximal temperature of the Universe is overestimated if one assumes the {\it instantaneous thermalization}~\cite{Harigaya:2013vwa, Mukaida:2015ria}. 
Second, 
the inflaton decay products, whose initial energy is of the same order as the inflaton mass, 
survive for a finite time until they lose their energy. 
These high-energy particles may be relevant for the DM production~\cite{Harigaya:2014waa}, 
particularly in the case where the cross section depends positively on the center of mass energy.

Non-thermal production of DM was extensively discussed in Refs.~\cite{Giudice:2000ex,Moroi:1994rs,Kawasaki:1995cy,Moroi:1999zb,Gelmini:2006pw} under the assumption of {\it instantaneous thermalization}. 
Ref.~\cite{Giudice:2000ex} discusses thermal production during the reheating era. 
Since there is an ambient plasma before the reheating completes, 
the DM can be produced from the scattering in the ambient plasma even if the reheating temperature is lower than the freeze-out temperature. 
Refs.~\cite{Moroi:1994rs,Kawasaki:1995cy,Moroi:1999zb,Gelmini:2006pw}
discuss the non-thermal production of DM from the inflaton (or moduli) decay. 
If the branching ratio is large enough, the number density of produced WIMPs is determined by the annihilation process. On the other hand, if the branching ratio is small, the number density is determined simply by the branching ratio and the decay temperature. 
There are also recent works in the literature~\cite{Garcia:2017tuj, Dudas:2017kfz, Drees:2017iod, Garcia:2018wtq, Allahverdi:2018aux, Kaneta:2019zgw}. 
These works except for Ref.~\cite{Garcia:2018wtq} do not take into account the finiteness of the thermalization time scale and estimate the DM production from the scattering of particles in the thermal plasma before the reheating is completed. 
Ref.~\cite{Garcia:2018wtq} takes into account the correct maximal temperature, which is much lower than the one obtained in the {\it instantaneous thermalization} approximation. 
Still, they omit the second effect we discussed above and may underestimate the DM production rate.

In this paper, we take into account the finiteness of thermalization time scale. 
There are three processes to produce DM during the thermalization and reheating era: 
(A) a collision between high-energy particles, 
(B) a collision between a high-energy particle and a particle in the ambient plasma, 
and (C) a collision between particles in the ambient plasma. 
We calculate DM abundance from these contributions and show the conditions to figure out the dominant process. 
We will see that the contribution (A) is dominant when the cross section of DM production depends positively and strongly on the center of mass energy as considered in Ref.~\cite{Garcia:2018wtq}. 
The contribution (B) is important when the cross section is mildly dependent on the center of mass energy and the reheating temperature is much lower than the DM mass.  This includes the case considered in Ref.~\cite{Harigaya:2014waa}.

\section{Thermalization era}

In this section, we briefly review the thermalization process 
of inflaton decay products, which is relevant to the computation of  the non-thermal  production rate of dark matter in the post-reheating era. 
Throughout this paper, we assume that the reheating proceeds via a perturbative decay of inflaton.
This is the case where the coupling between inflaton and radiation is tiny, \textit{e.g.}, Planck-suppressed operators, because the resonance tends to be shut off immediately due to the cosmic expansion.

As we explained in the introduction, 
the thermalization of high-energy particles proceeds via the splitting into lower energy particles.
This process is the bottleneck of the thermalization if the energy of the high-energy particles, $\sim m_\phi$, is larger than the would-be temperature of radiation after thermalization, \textit{i.e.}, $m_\phi > T$.
The rate of the splitting process 
is suppressed by the destructive interference effect between the parent particle 
and the daughter particle, which is known as the LPM effect~\cite{Landau:1953um, Migdal:1956tc, Gyulassy:1993hr, Arnold:2001ba, Arnold:2001ms, Arnold:2002ja, Besak:2010fb}. 
Once the splitting occurs, the subsequent process 
does not occur until a phase factor, $kx$, varies significantly ($\sim 1$), 
where $k$ is the 4-momentum of a daughter particle 
and $x$ is the position of the parent particle. 
Denoting the emission angle of the daughter particle as $\theta$ ($= k_\perp / k$, 
where $k_\perp$ is the perpendicular momentum), 
we obtain $1 \lesssim k x \sim t k \theta^2 \sim t k_\perp^2 / k$. 
Here we have assumed that all the particles are relativistic, $k^0 \simeq k$.
This inequality puts a minimum time so as to emit particles, known as a formation time, $t_\text{form}$.

Now we need to estimate how the squared perpendicular momentum of the daughter particle, $k_\perp^2$, evolves in the medium.\footnote{
	If there is no ambient plasma, $k_\perp$ is kept intact. Thus the emitting occurs only for $\theta > 1 / (kt)^{1/2}$ where we expect DGLAP vacuum shower dominates the splittings. See Ref.~\cite{Kurkela:2014tla} for the classification of splittings processes.
}
When the daughter particle is scattered by elastic scattering processes, 
it evolves as a random walk as~\cite{Kurkela:2011ti}
\begin{align}
 &(k_\perp)^2 \sim \hat{q}_{\rm el} t, \label{eq:random}
 \\
 &\hat{q}_{\rm el} \sim \int \dd^2 q_\perp \frac{\del \Gamma_{\rm el}}{\del q_\perp^2} q_\perp^2 \sim \alpha^2 \int \dd^3 k\, f (k) \sim \alpha^2 T^3, 
 \label{eq:diff_cnst}
\end{align}
where we used the elastic scattering rate
\begin{align}
 \frac{\del \Gamma_{\rm el}}{\del q_\perp^2} \sim \frac{\alpha^2}{q_\perp^2 (q_\perp^2 + \alpha T^2)} 
 \int \dd^3 k\, f (k). 
\end{align}
Here, $f (k)$ denotes a phase space distribution 
and $\alpha$ generically represents fine-structure constants for the SM gauge interactions. 
In the last similarity in Eq.~\eqref{eq:diff_cnst}, we assume that the number density is dominated by the thermalized sector.\footnote{
	In fact one may show that this is the case for $T < T_\text{max}$ which we will focus on throughout this paper. See Refs.~\cite{Harigaya:2013vwa, Mukaida:2015ria}.
}
Inserting these Eqs.~\eqref{eq:random} and \eqref{eq:diff_cnst} into $1 \lesssim t k_\perp^2 / k$, one may estimate the formation time
\begin{align}
	t_\text{form} (k) := \left( \frac{k}{\hat{q}_\text{el}} \right)^{1/2},
\end{align}
which is the minimum time to emit particles.
If this is longer than the time scale of elastic scatterings, $1 / \Gamma_\text{el}$, the splitting rate is suppressed.
As a result, we obtain 
\begin{align}
 \Gamma_{\rm split}(k) \sim \alpha \min \left[ \Gamma_\text{el}, \frac{1}{t_\text{form}(k)} \right].
\end{align}
In the following we focus on the latter rate \textit{i.e.}, the LPM suppressed rate, dominates the splitting, which implies $k \gtrsim k_\text{LPM} \equiv \hat{q}_\text{el} / \Gamma_\text{el}^2 \sim T$.
Note that the splitting rate via the LPM effect does not depend on the energy of the parent particle in non-Abelian gauge theories.
The energy loss rate for the parent particle is dominated by the largest possible energy for the daughter particle $k$, 
which is equal to half of the parent energy. 
This means that 
the high-energy particles continuously lose their individual energy via the splitting into high-energy particles.

When $\Gamma_{\rm split} (m_\phi) \gtrsim H$ is satisfied, 
the inflaton decay products lose their energy within a Hubble time 
and are thermalized soon after they are produced. 
The threshold of this regime, which we denote as $t_{\rm max}$, is the time when the temperature reaches its maximal value after inflation. 
The corresponding temperature turns out to be~\cite{Harigaya:2013vwa, Mukaida:2015ria}
\begin{align} 
 T_{\rm max} \sim \alpha^{4/5} m_\phi \lmk \frac{\Gamma_\phi \Mpl^2}{m_\phi^3} \rmk^{2/5}, 
 \label{Tmax}
\end{align}
where $\Gamma_\phi$ is the decay rate of the inflaton. 
After this happens $t \gtrsim t_\text{max}$, the energy and number density of radiation is dominated by the thermalized sector.
The estimation of the maximum temperature is valid only when $T_{\rm max} < m_\phi$ which we assume in the following. Our estimation of the DM abundance is valid without this assumption if the production is dominated at the temperature much below $T_{\rm max}$. 

High-energy particles are continuously produced from the inflaton decay 
until the Hubble parameter decreases to the decay rate: $\Gamma_\phi \simeq H$. 
This is the time when the reheating is completed and the Universe becomes to be dominated by thermal radiation. We denote the time as $t_{\text{RH}}$.
From $t_{\rm max}$ to $t_{\text{RH}}$, 
the temperature of the Universe decreases according to 
\begin{align}
 \rho_r \sim T^4 \sim H \Gamma_\phi \Mpl^2, 
\end{align}
The temperature at $t = t_{\text{RH}}$, which is commonly called as the reheating temperature, is given by 
\begin{align}
 T_{\text{RH}} \sim \sqrt{\Gamma_\phi \Mpl}. 
 \label{TRH}
\end{align}

\begin{figure}[t] 
\centering
\includegraphics[width=0.6 \textwidth]{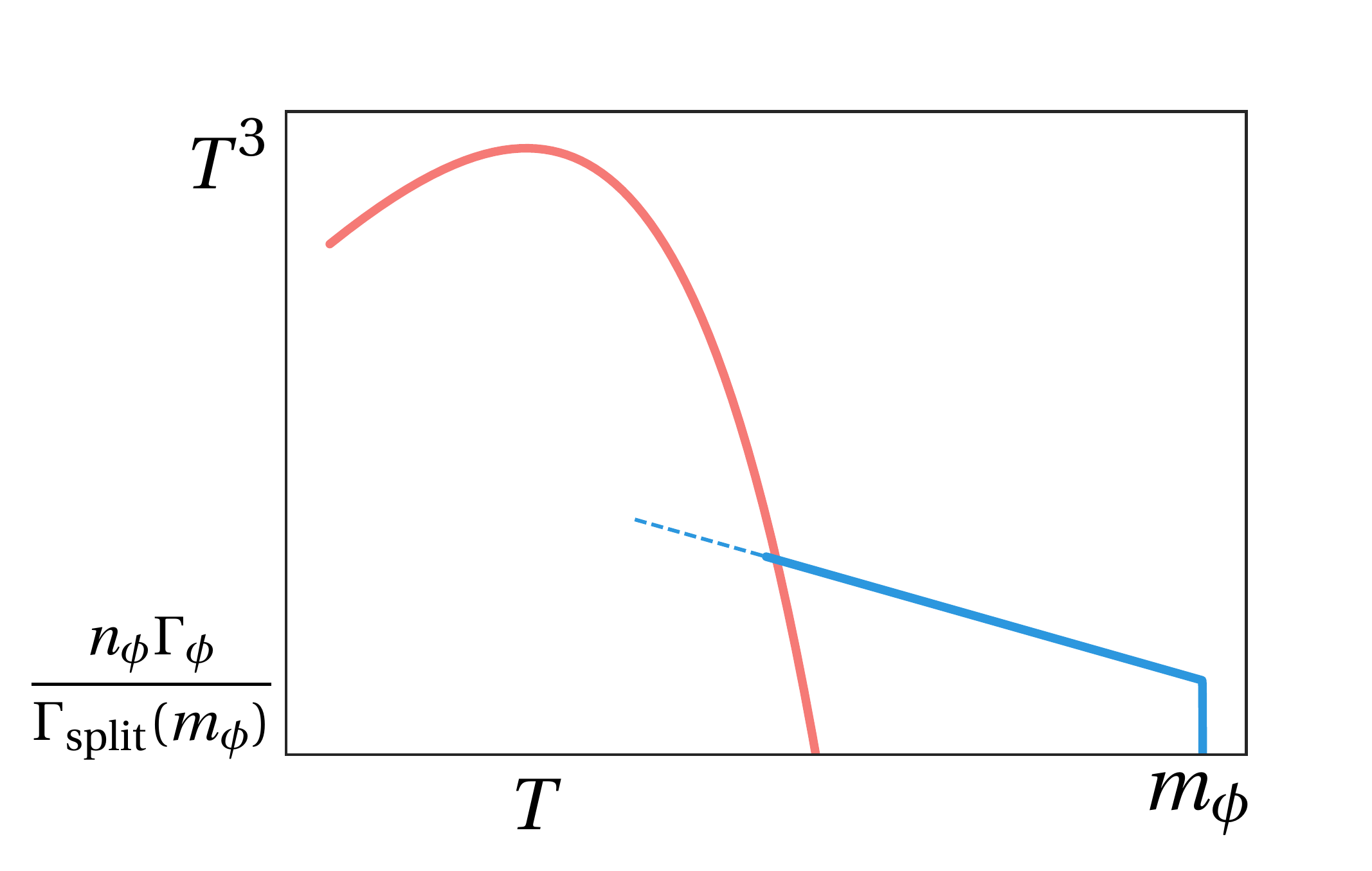} 
\caption{A schematic plot of $n(E)$ as a function of $E$.
While the red curve shows the thermal distribution given in Eq.~\eqref{eq:thermal},
the blue curve shows the high-energy tail given in Eq.~\eqref{eq:tail}.}
\label{fig:spectrum}
\end{figure}

Now we are in a position to discuss a tail of cascading particles which can play an essential role for the non-thermal DM production.
Let us emphasize that the high-energy particles are continuously produced from the inflaton decay during the reheating. 
The time scale of the thermalization of these high-energy particles is given by the splitting rate $1/\Gamma_{\rm split} (m_\phi)$, which is finite.
As a result, although this process is completed within a Hubble time for $T < T_\text{max}$, there remains a tail of cascading particles because the inflaton continuously sources the high-energy particles.
The spectrum of these cascading particles can be estimated as follows.
For later convenience, we define the number density with energy $E$ denoted by $n(E)$ as follows:
\begin{align}
	n = \int \dd^3 k f(k) =: \int \dd \log E \, n (E).
\end{align}
The number density of high-energy particles right after the inflaton decay is $n_\phi \Gamma_\phi / H$.
These primary particles with energy $\sim m_\phi$ cascade down to some energy $E$ by splittings.
The energy conservation naively implies $n(E) \sim (m_\phi / E) n_\phi \Gamma_\phi / H$.
However this population never stays the same energy $E$ for the time interval $H^{-1}$, rather they further cascade within $1 / \Gamma_\text{split}(E)$.
By taking this effect into account, we can estimate the spectrum~\cite{Harigaya:2014waa} (see also \cite{Kurkela:2014tla}):
\begin{align}
	n (E) \sim 	
	\frac{m_\phi}{E} n_\phi \frac{\Gamma_\phi}{H} \frac{H}{\Gamma_\text{split}(E)}
		\sim n_\phi \frac{\Gamma_\phi}{\Gamma_\text{split} (m_\phi)}
		\left( \frac{m_\phi}{E} \right)^{1/2}
		\label{eq:tail}
\end{align}
for $E > T$ (for a detailed calculation, see Appendix in Ref.~\cite{Harigaya:2014waa}). Here $n_\phi$ ($\simeq 3 H^2 \Mpl^2 / m_\phi$) is the number density of the inflaton.
The characteristic power-law of $E^{-1/2}$ shows the LPM suppression.
After cascading down to $T$, the daughter particles get thermalized by their own interactions which results in
\begin{align}
	n (E) \sim n_\text{eq} (E)
	\label{eq:thermal}
\end{align}
for $E < T$. Here $n_\text{eq} (E)$ denotes the thermal distribution with the temperature $T$.

The whole spectrum, $n(E)$, is shown in Fig.~\ref{fig:spectrum}. 
One can see that taking $\Gamma_{\rm split} \to \infty$ (i.e., assuming the {\it instantaneous thermalization}), results in no high energy population. We also note that the high energy tail is inevitable even if $\Gamma_{\rm split} > H$ as long as $\Gamma_{\rm split}$ is finite. 
Note again that the high-energy tail in the spectrum never dominates the energy and number density of radiation.
Nevertheless this tail can dominate the production of heavy particles such as DM because its energy is much larger than the temperature for $m_\phi \gg T$.
As we will see, the scattering involving particles in the high-energy tail is actually the dominant process 
when the cross section of DM production process positively depends on 
the center of mass energy and/or the reheating temperature is smaller than the DM mass.

\section{DM production during the thermalization era}

Now we shall take into account the production of DM 
from those particles. We assume that DM is produced from a pair of SM particles.
We also assume that the direct decay of the inflaton into DM is negligible. (This process tends to be non-negligible when DM is charged under the SM gauge group, as DM is necessarily produced by the showering process~\cite{Kurata:2012nf,Harigaya:2016vda}.)

We parametrize the cross section for the DM production as 
\begin{align}
 \sigma_{\rm DM} (E_{\rm CM}) = \frac{E_{\rm CM}^{n}}{M^{n+2}}, 
 \label{sigma}
\end{align}
where $E_{\rm CM}$ is the center of mass energy 
and 
$M$ is a parameter with a mass dimension $1$. 
This kind of cross section comes from a higher dimensional operator 
with a cutoff of order $M$~\cite{Mambrini:2013iaa, Mambrini:2015vna, Soni:2016gzf, Benakli:2017whb, Dudas:2017rpa, Kang:2019izi}, 
where we absorb a coupling constant in the definition of $M$ 
without loss of generality. 
The operator comes from an interaction mediated by a mediator of a mass of order $M$. 
Note that the dependence of the cross section on $E_{\rm CM}$ is modified for $E_{\rm CM} \gtrsim M$ to recover the unitarity. 
We should replace it by 
$1/E_{\rm DM}^2$ for $E_{\rm CM} \gtrsim M$. 
This overestimates the production rate for $E_{\rm CM} >M$, but practically the production is dominated at $E_{\rm CM} \sim M$. In the following discussion, we assume $E_{\rm CM} \lesssim M$ 
for simplicity. The estimation in the opposite case is obtained by taking $n = 0$
in the following calculation.

Suppose that the DM is produced from a collision of a particle with energy $E_1$ and a particle with energy $E_2$. 
The rate of DM production for the former particle is then given by 
\begin{align}
 \Gamma_{\rm DM} (E_1,E_2) \sim \frac{(E_1 E_2)^{n/2}}{M^{n+2}} n (E_2). 
\end{align}
The total number density of DM generated per a Hubble time is obtained by performing integrals over the energies $E_{1,2},$
\begin{align}
	n_\text{DM} = \int \dd \log E_1 \dd \log E_2 \, n_\text{DM} (E_1, E_2),
	\label{eq:tot_DM_num}
\end{align}
where
\begin{align}
	n_\text{DM} (E_1, E_2) \sim \frac{\Gamma_\text{DM} (E_1, E_2)}{H} n(E_1).
\end{align}

The rest of this paper is devoted to the determination of  the dominant contribution of $n_\text{DM} (E_1, E_2)$ in the integral \eqref{eq:tot_DM_num} by inserting $n(E)$ in Eqs.~\eqref{eq:tail} and \eqref{eq:thermal}. In the following calculation, 
we assume
$m_{\rm DM} < m_\phi, T_{\rm max}$. 
We also assume that the annihilation of produced DM is negligible.

\subsection{(A) Collisions of particles with $E_1$ and $E_2$ ($>T$)}

In this subsection, we calculate the DM abundance produced from 
the collision between the high-energy particles with $T < E_1,E_2 \lesssim m_\phi$. 
The DM number density produced during one Hubble time in the 
thermalization era at the temperature of $T$ ($T_{\text{RH}} < T < T_{\rm max}$) is given by 
\begin{align}
 \left. \frac{n_{\rm DM}^{(\rm NT)} (E_1, E_2) }{s} \right\vert_T
 \sim& \frac{\Gamma_{\rm DM}(E_1,E_2)}{H(T)} \frac{n(E_1) }{s}, 
 \label{formula}
 \\
 \sim& 
 \frac{m_\phi^{n-1} T_{\text{RH}}^3 T}{\alpha^4 \Mpl M^{n+2}} 
 \lmk \frac{E_1 E_2}{m_\phi^2} \rmk^{(n-1)/2}.
 \label{formula2}
\end{align}

The subscript (NT) implies that the production involves the high-energy tail, \textit{i.e.}, $T < E_1$ or $T < E_2$.
Here we normalize the number density by the effective ``entropy density" $1/s$.
We define $s$ by $4 \rho_\phi / 3 T_{\text{RH}}$ during the reheating era so 
that $s$ is proportional to $1/a^3$ and is equal to the entropy density at $T = T_{\text{RH}}$.
In other words, the entropy production via the inflaton decay is already factored out in its definition.
By using this definition of $s$, the combination of $n_{\rm DM} / s$ is constant in time if there is no additional source of DM nor the entropy.

Equation~(\ref{formula}) is the formula derived in Ref.~\cite{Harigaya:2014waa}. 
This contribution is missed 
if one takes a limit of $\Gamma_{\rm split} \to \infty$, which corresponds to the limit of {\it instantaneous thermalization}. 
From \eq{formula2}, 
we can see that the maximal values,
$E_1 \sim m_\phi$ and $E_2  \sim m_\phi$, dominate the integral \eqref{eq:tot_DM_num} for $n > 1$. 
For $n < 1$ the lowest possible values, $E_1 = E_2 = m_\text{DM}$, are dominant. 
If we set $E_1 = E_2 \sim m_\phi$ or $m_{\rm DM}$, 
the resulting DM abundance is proportional to $T$. 
Therefore, 
DM with energy of order $m_\phi$ is produced dominantly at the time when 
the temperature reaches its maximal value $T_{\rm max}$. 
Then we obtain 
\begin{align}
 &&
 \frac{n_{\rm DM}^{(\rm NT)} (E_1, E_2) }{s} 
 \sim 
 \frac{m_\phi^{n-6/5} T_{\text{RH}}^{19/5}}{\alpha^{16/5} \Mpl^{3/5} M^{n+2}}
  \lmk \frac{E_1 E_2}{m_\phi^2} \rmk^{(n-1)/2},
\end{align}
with the dominant contribution coming from 
\begin{align}
 E_1 E_2 \to 
 \begin{cases}
 m_{\rm DM}^2 
 &\qquad \text{for} \quad n < 1
 \\
 m_\phi^2 &\qquad \text{for} \quad n > 1. 
 \end{cases}
\end{align}
Here,  $E_1 E_2$ must be larger than $T_{\rm max}^2$ by assumption, which requires $T_{\rm max} \lesssim m_{\rm DM}$ for $n< 1$. Otherwise the contribution B or C is dominant. 
The result in the latter case ($n > 1$) is obtained in Ref.~\cite{Garcia:2018wtq},
although the contributions from $T<T_{\rm max}$ is not discussed and it is not clear if they are subdominant.
For $n=1$, where the dependence of the DM abundance on $E_{1,2}$ vanishes, the integration over $E_{1,2}$ yields logarithmic factors.
This is also true for the other contributions B and C when the dependence on $E_{1,2}$ and/or $T$ vanishes. 
In this paper, we neglect these factors for simplicity.

Finally, we comment that the contribution coming from the time before the temperature reaches the maximal value (\textit{i.e.}, the time when $H > \Gamma_{\rm split}$) is always subdominant. In this case, the number of high-energy particles is given by \eq{eq:tail} with the replacement of $1/\Gamma_{\rm split} (E) \to 1/H$. The factor of $1/H^2$ from $n (E_1) n (E_2)$ is larger for a later epoch, so that the contribution is maximized when the temperature reaches the maximal value.

\subsection{(B) Collisions of particles with $E_1$ ($>T$) and $T$}

In this subsection, we calculate the DM abundance produced from 
the interaction between the high-energy particle with energy of $m_\phi \gtrsim E_1 > T$ and the thermal plasma $T > E_2$. 
We assume that $T_{\rm max} m_{\phi} > m_{\rm DM}$. Otherwise the produced DM abundance is exponentially suppressed. 
The result is given by 
\begin{align}
 &\left. \frac{n_{\rm DM}^{(\rm NT)} (E_1, T) }{s}  \right\vert_T
 \nn
 \sim& 
 \frac{m_\phi^{n-3} T_{\text{RH}}^5}{\alpha^2 M^{n+2}} 
 \lmk \frac{E_1}{m_\phi} \rmk^{(n-1)/2} 
 \lmk \frac{T}{m_\phi} \rmk^{(n-5)/2}. 
 \label{formula3}
\end{align}
Here we have already utilized the fact that the thermal population is dominated by $E_2 \sim T$.
The dominant contribution comes from 
\begin{align}
 E_1 \to 
 \begin{cases}
 {\rm Min} \lkk m_{\rm DM}^2 / T_{\text{RH}}, \ m_\phi \rkk
 &\qquad \text{for} \quad n < 1
 \\
 m_\phi &\qquad \text{for} \quad n > 1,
 \end{cases}
\end{align}
and 
\begin{align}
 T \to 
 \begin{cases}
 {\rm Max} \lkk T_{\text{RH}}, \ m_{\rm DM}^2 / m_\phi \rkk
 &\qquad \text{for}  \quad n < 5
 \\
 T_{\rm max} &\qquad \text{for} \quad n > 5,
 \end{cases}
\end{align}
Here, we should note that $E_1$ must be larger than $T$ by assumption, 
which may be violated for $m_{\rm DM} \lesssim T_{\text{RH}}$ and $n <1$. 
Otherwise the contribution C is dominant.

The ratio between the contribution A and B is given by 
\begin{align}
 \frac{n_{\rm DM}^{(\rm NT)} (m_\phi, m_\phi)}{ n_{\rm DM}^{(\rm NT)}(m_\phi, T)} 
 \sim 
  \begin{cases}
 \lmk \frac{T_{\rm RH}}{T_{\rm max}} \rmk^{3/2} 
 \lmk \frac{T_{\rm RH}}{m_\phi} \rmk^{1/2} 
 &~~\text{for}  \quad n < 1
 \vspace{0.2cm}
 \\
  \lmk \frac{T_{\rm RH}}{T_{\rm max}} \rmk^{3/2} 
 \lmk \frac{T_{\rm RH}}{m_\phi} \rmk^{(2-n)/2} 
 &~~\text{for} \quad 1< n < 5,
 \vspace{0.2cm}
 \\
 \lmk \frac{m_\phi}{T_{\rm max}} \rmk^{(n-2)/2} 
 &~~\text{for} \quad 5 < n,
 \end{cases}
 \label{ratio1}
\end{align}
where we assumed $m_{\rm DM}^2 / T_{\rm RH} < m_\phi$. 
Under the assumption of $T_{\rm max} \lesssim m_\phi$, which is the case for $\Gamma_\phi \lesssim m_\phi^3 / \Mpl^2$,
the contribution A is dominant for $n >5$. 
One can check that the contribution B is dominant 
for $n<2$ because $T_{\text{RH}} < T_{\rm max}$.

\subsection{(C) Collisions of particles in the thermal plasma}

Finally, we calculate the DM abundance produced from the thermal plasma. We assume that $T_{\rm max} > m_{\rm DM}$. Otherwise the contribution is exponentially suppressed.

The DM number density produced from the 
thermal plasma at $T$ ($\in (T_{\text{RH}}, T_{\rm max})$) 
is calculated in Ref.~\cite{Garcia:2017tuj}.
The result is given by 
\begin{align}
 \left. \frac{n_{\rm DM}^{(\rm T)}(T,T)}{s}  \right\vert_T 
 \sim \frac{T^{n-6} T_{\text{RH}}^7 \Mpl}{M^{n+2}}, 
\end{align}
where we used Eqs.~\eqref{TRH} and \eqref{eq:thermal}. Here we implicitly assumed that $m_{\rm DM} < T$ 
so that DM can be produced from the thermal plasma.
The main contribution comes from 
\begin{align}
 T \to 
 \begin{cases}
 {\rm Max} \lkk T_{\text{RH}}, \ m_{\rm DM} \rkk
 &\qquad \text{for} \quad n < 6
 \\
 T_{\rm max} &\qquad \text{for} \quad n > 6. 
 \end{cases}.
 \label{T3}
\end{align}
When $T_{\rm RH} > m_{\rm DM}$ and $n < -1$, the production dominantly occurs after reheating at $T \sim m_{\rm DM}$, with the abundance given by
\begin{align}
 \left. \frac{n_{\rm DM}^{(\rm T)}(T,T)}{s}  \right\vert_{T\sim m_{\rm DM}} \sim \frac{m_{\rm DM}^{n+1} \Mpl}{M^{n+2}}.
\end{align}

The ratio between the contributions A and C is given by 
\begin{align}
 \frac{n_{\rm DM}^{(\rm NT)} (m_\phi, m_\phi)}{ n_{\rm DM}^{(\rm T)}} 
 \sim 
 \lmk \frac{m_\phi }{\alpha^2 \Mpl} \rmk^{8/5}
 \lmk \frac{m_\phi}{T_{\text{RH}}} \rmk^{n-14/5}, 
 \label{ratio2}
\end{align}
for $1<n < 6$ and $m_{\rm DM} < T_{\text{RH}}$. 
This implies that 
the contribution C is less important for $n \gtrsim 14/5 \simeq 3$.

We compare the contributions B and C 
for $n = 0,$ $2$ 
and find that the contribution C dominates when 
\begin{align}
\frac{m_\text{DM}^{5/4}}{\alpha^{1/2}   \Mpl^{1/4}}  \lesssim T_{\text{RH}} 
 &\qquad \text{for} \quad n = 0, 
 \label{th1}
 \\
 \lmk \frac{m_{\rm DM}^8 m_\phi}{\alpha^4 \Mpl^2} \rmk^{1/7} \lesssim T_{\text{RH}}, 
 &\qquad \text{for} \quad n = 2, 
 \label{th2}
\end{align}
where we assume a large $m_\phi$.

\begin{figure}
\centering
\includegraphics[width=0.49 \textwidth]{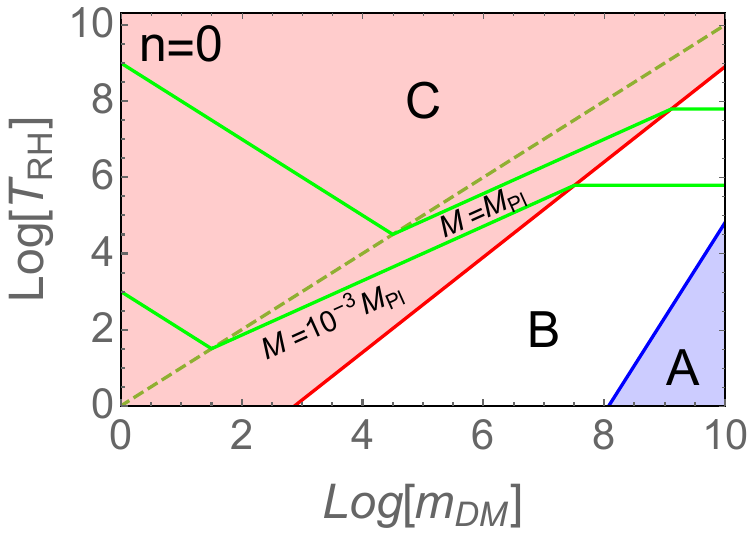} 
\includegraphics[width=0.49 \textwidth]{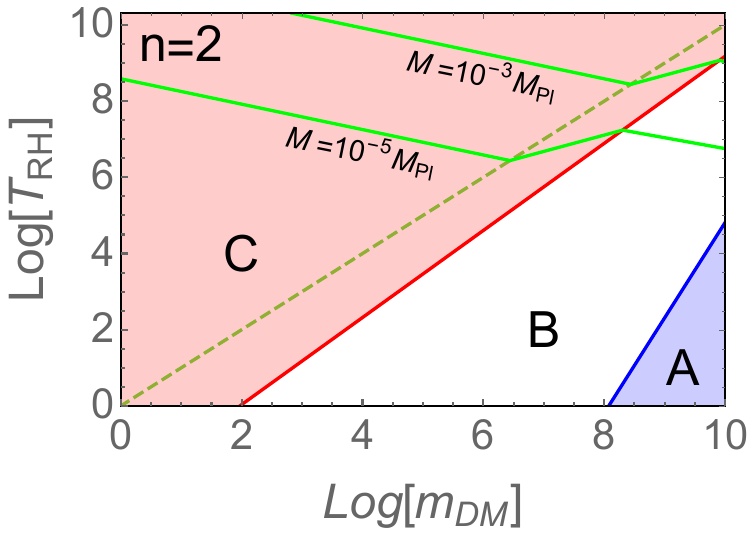} 
\caption{
Dominant contribution for the production of DM during the reheating epoch 
in $T_{\text{RH}} - m_{\rm DM}$ space. 
The blue, white, red regions represent parameter spaces that the contribution A, B, and C 
is dominant, respectively. 
We set $n = 0$ (left panel) or $n=2$ (right panel). 
Each green line represents parameters where we can explain the observed amount of DM with $M$ being the value shown in the figure. The dashed line represents $T_{\rm RH} = m_{\rm DM}$. 
}
\label{fig}
\end{figure}

Fig.~\ref{fig}
shows which contribution is dominant for $n = 0$ and $2$. 
Here we take $\alpha = 0.01$ and $m_\phi = 10^{13} \GeV$. 
We can see that the contribution $B$ dominates over $A$ for a low-reheating temperature, $T_{\text{RH}} \ll m_{\rm DM}$, as we discussed in Ref.~\cite{Harigaya:2014waa}. 
In the figure, we show the parameter region which explains the observed amount of DM 
($m_{\rm DM} n_{\rm DM}/s \simeq 0.4 \ {\rm eV}$) with $M = 10^{-3} \Mpl$ or $\Mpl$ for $n=0$ and $M = 10^{-5} \Mpl$ or $10^{-3}\Mpl$ for $n=2$. 
For a guide to the eye, we show a dashed line representing $T_{\rm RH} = m_{\rm DM}$, 
which is the threshold determining the dominant contribution for $T$ in \eq{T3}.

We have checked that 
the contribution A is almost always dominant for $n \gtrsim 4$. 
The contribution C can be dominant for $n \simeq 3$, where the threshold is given by \eq{ratio2}. 
For $n \lesssim 2$, 
the contributions B or C are dominant depending on $m_{\rm DM}$ and $T_{\text{RH}}$, 
where the threshold is given by \eq{th1} or \eq{th2}. 
The contribution A is dominant for $n \lesssim 2$ only in the case where $m_{\rm DM}^2 / T_{\text{RH}} \gg m_\phi$ as one can see at the right-bottom corner in Fig.~\ref{fig}.

\section{Conclusions and discussion}

We have calculated 
the abundance of DM produced during the thermalization and reheating era after inflation, 
taking into account the finiteness of thermalization time for high-energy particles from the inflaton decay. 
The thermalization proceeds via splitting into high-energy particles, 
whose rate is suppressed by the LPM effect. 
There are two important effects that are missed in the {\it instantaneous thermalization} approximation. 
First, the temperature of the Universe does not reach its maximal value until the thermalization is completed. This means that the maximal temperature is overestimated in the {\it instantaneous thermalization} approximation. 
Second, 
the high-energy particles that are continuously produced from inflaton decay 
can produce DM before they lose their energy to be thermalized. This process is important when the DM production cross section is positively dependent on the center of mass energy and/or when the DM mass is larger than the reheating temperature.

There are three processes to produce DM during the reheating era: 
A) a collision between high-energy particles,
B) a collision between a high-energy particle and a particle in the ambient plasma, 
and C) collision between particles in the ambient plasma. 
The first one is discussed in Ref.~\cite{Garcia:2018wtq} though the second contribution is omitted.
The second one is discussed in Ref.~\cite{Harigaya:2014waa} only for the case of $n = 0$. 
In this paper, we have generalized the discussion in these papers and 
have determined the conditions for each contribution to be dominant. 
We have checked that the first contribution is dominant for the case of $n \gtrsim 3$, 
where $n$ is the power of the dependence of the DM production cross section on the center of mass energy. 
The second contribution is dominant for the case where the reheating temperature is much lower than the DM mass. 
These results are consistent with the results in Refs.~\cite{Harigaya:2014waa, Garcia:2018wtq} when overlapped. The resulting DM abundance is many orders of magnitude larger than the one calculated in the {\it instantaneous thermalization} approximation in most cases.

\section*{Acknowledgements}
The work was supported in part by the DoE grant DE-SC0009988 (K.H.).
KH thanks Raymond Co and Aaron Pierce for fruitful discussion during which an error in Eq.~(\ref{th1}) of the previous version of the paper was spotted.

\small
\bibliography{reference}

\providecommand{\href}[2]{#2}\begingroup\raggedright\begin{thebibliography}{10}

\bibitem{Lee:1977ua}
B.~W. Lee and S.~Weinberg, ``{Cosmological Lower Bound on Heavy Neutrino
  Masses},''
\href{http://dx.doi.org/10.1103/PhysRevLett.39.165}{{\em Phys. Rev. Lett.}
  {\bfseries 39} (1977) 165--168}.

\bibitem{Harigaya:2014waa}
K.~Harigaya, M.~Kawasaki, K.~Mukaida, and M.~Yamada, ``{Dark Matter Production
  in Late Time Reheating},''
  \href{http://dx.doi.org/10.1103/PhysRevD.89.083532}{{\em Phys. Rev.}
  {\bfseries D89} no.~8, (2014) 083532},
\href{http://arxiv.org/abs/1402.2846}{{\ttfamily arXiv:1402.2846 [hep-ph]}}.

\bibitem{Starobinsky:1980te}
A.~A. Starobinsky, ``{A New Type of Isotropic Cosmological Models Without
  Singularity},''
\href{http://dx.doi.org/10.1016/0370-2693(80)90670-X}{{\em Phys. Lett.}
  {\bfseries B91} (1980) 99--102}.

\bibitem{Traschen:1990sw}
J.~H. Traschen and R.~H. Brandenberger, ``{Particle Production During
  Out-of-equilibrium Phase Transitions},''
\href{http://dx.doi.org/10.1103/PhysRevD.42.2491}{{\em Phys. Rev.} {\bfseries
  D42} (1990) 2491--2504}.

\bibitem{Kofman:1994rk}
L.~Kofman, A.~D. Linde, and A.~A. Starobinsky, ``{Reheating after inflation},''
  \href{http://dx.doi.org/10.1103/PhysRevLett.73.3195}{{\em Phys. Rev. Lett.}
  {\bfseries 73} (1994) 3195--3198},
\href{http://arxiv.org/abs/hep-th/9405187}{{\ttfamily arXiv:hep-th/9405187
  [hep-th]}}.

\bibitem{Shtanov:1994ce}
Y.~Shtanov, J.~H. Traschen, and R.~H. Brandenberger, ``{Universe reheating
  after inflation},'' \href{http://dx.doi.org/10.1103/PhysRevD.51.5438}{{\em
  Phys. Rev.} {\bfseries D51} (1995) 5438--5455},
\href{http://arxiv.org/abs/hep-ph/9407247}{{\ttfamily arXiv:hep-ph/9407247
  [hep-ph]}}.

\bibitem{Kofman:1997yn}
L.~Kofman, A.~D. Linde, and A.~A. Starobinsky, ``{Towards the theory of
  reheating after inflation},''
  \href{http://dx.doi.org/10.1103/PhysRevD.56.3258}{{\em Phys. Rev.} {\bfseries
  D56} (1997) 3258--3295},
\href{http://arxiv.org/abs/hep-ph/9704452}{{\ttfamily arXiv:hep-ph/9704452
  [hep-ph]}}.

\bibitem{Felder:1998vq}
G.~N. Felder, L.~Kofman, and A.~D. Linde, ``{Instant preheating},''
  \href{http://dx.doi.org/10.1103/PhysRevD.59.123523}{{\em Phys. Rev.}
  {\bfseries D59} (1999) 123523},
\href{http://arxiv.org/abs/hep-ph/9812289}{{\ttfamily arXiv:hep-ph/9812289
  [hep-ph]}}.

\bibitem{Harigaya:2013vwa}
K.~Harigaya and K.~Mukaida, ``{Thermalization after/during Reheating},''
  \href{http://dx.doi.org/10.1007/JHEP05(2014)006}{{\em JHEP} {\bfseries 05}
  (2014) 006},
\href{http://arxiv.org/abs/1312.3097}{{\ttfamily arXiv:1312.3097 [hep-ph]}}.

\bibitem{Mukaida:2015ria}
K.~Mukaida and M.~Yamada, ``{Thermalization Process after Inflation and
  Effective Potential of Scalar Field},''
  \href{http://dx.doi.org/10.1088/1475-7516/2016/02/003}{{\em JCAP} {\bfseries
  1602} no.~02, (2016) 003},
\href{http://arxiv.org/abs/1506.07661}{{\ttfamily arXiv:1506.07661 [hep-ph]}}.

\bibitem{Davidson:2000er}
S.~Davidson and S.~Sarkar, ``{Thermalization after inflation},''
  \href{http://dx.doi.org/10.1088/1126-6708/2000/11/012}{{\em JHEP} {\bfseries
  11} (2000) 012},
\href{http://arxiv.org/abs/hep-ph/0009078}{{\ttfamily arXiv:hep-ph/0009078
  [hep-ph]}}.

\bibitem{Jaikumar:2002iq}
P.~Jaikumar and A.~Mazumdar, ``{Postinflationary thermalization and
  hadronization: QCD based approach},''
  \href{http://dx.doi.org/10.1016/j.nuclphysb.2004.02.015}{{\em Nucl. Phys.}
  {\bfseries B683} (2004) 264--276},
\href{http://arxiv.org/abs/hep-ph/0212265}{{\ttfamily arXiv:hep-ph/0212265
  [hep-ph]}}.

\bibitem{Arnold:2002zm}
P.~B. Arnold, G.~D. Moore, and L.~G. Yaffe, ``{Effective kinetic theory for
  high temperature gauge theories},''
  \href{http://dx.doi.org/10.1088/1126-6708/2003/01/030}{{\em JHEP} {\bfseries
  01} (2003) 030},
\href{http://arxiv.org/abs/hep-ph/0209353}{{\ttfamily arXiv:hep-ph/0209353
  [hep-ph]}}.

\bibitem{Kurkela:2011ti}
A.~Kurkela and G.~D. Moore, ``{Thermalization in Weakly Coupled Nonabelian
  Plasmas},'' \href{http://dx.doi.org/10.1007/JHEP12(2011)044}{{\em JHEP}
  {\bfseries 12} (2011) 044},
\href{http://arxiv.org/abs/1107.5050}{{\ttfamily arXiv:1107.5050 [hep-ph]}}.

\bibitem{York:2014wja}
M.~C. Abraao~York, A.~Kurkela, E.~Lu, and G.~D. Moore, ``{UV cascade in
  classical Yang-Mills theory via kinetic theory},''
  \href{http://dx.doi.org/10.1103/PhysRevD.89.074036}{{\em Phys. Rev.}
  {\bfseries D89} no.~7, (2014) 074036},
\href{http://arxiv.org/abs/1401.3751}{{\ttfamily arXiv:1401.3751 [hep-ph]}}.

\bibitem{Kurkela:2014tea}
A.~Kurkela and E.~Lu, ``{Approach to Equilibrium in Weakly Coupled Non-Abelian
  Plasmas},'' \href{http://dx.doi.org/10.1103/PhysRevLett.113.182301}{{\em
  Phys. Rev. Lett.} {\bfseries 113} no.~18, (2014) 182301},
\href{http://arxiv.org/abs/1405.6318}{{\ttfamily arXiv:1405.6318 [hep-ph]}}.

\bibitem{Kurkela:2014tla}
A.~Kurkela and U.~A. Wiedemann, ``{Picturing perturbative parton cascades in
  QCD matter},'' \href{http://dx.doi.org/10.1016/j.physletb.2014.11.054}{{\em
  Phys. Lett.} {\bfseries B740} (2015) 172--178},
\href{http://arxiv.org/abs/1407.0293}{{\ttfamily arXiv:1407.0293 [hep-ph]}}.

\bibitem{Drewes:2010pf}
M.~Drewes, ``{On the Role of Quasiparticles and thermal Masses in
  Nonequilibrium Processes in a Plasma},''
\href{http://arxiv.org/abs/1012.5380}{{\ttfamily arXiv:1012.5380 [hep-th]}}.

\bibitem{Mukaida:2012qn}
K.~Mukaida and K.~Nakayama, ``{Dynamics of oscillating scalar field in thermal
  environment},'' \href{http://dx.doi.org/10.1088/1475-7516/2013/01/017}{{\em
  JCAP} {\bfseries 1301} (2013) 017},
\href{http://arxiv.org/abs/1208.3399}{{\ttfamily arXiv:1208.3399 [hep-ph]}}.

\bibitem{Mukaida:2012bz}
K.~Mukaida and K.~Nakayama, ``{Dissipative Effects on Reheating after
  Inflation},'' \href{http://dx.doi.org/10.1088/1475-7516/2013/03/002}{{\em
  JCAP} {\bfseries 1303} (2013) 002},
\href{http://arxiv.org/abs/1212.4985}{{\ttfamily arXiv:1212.4985 [hep-ph]}}.

\bibitem{Drewes:2013iaa}
M.~Drewes and J.~U. Kang, ``{The Kinematics of Cosmic Reheating},''
  \href{http://dx.doi.org/10.1016/j.nuclphysb.2013.07.009,
  10.1016/j.nuclphysb.2014.09.008}{{\em Nucl. Phys.} {\bfseries B875} (2013)
  315--350}, \href{http://arxiv.org/abs/1305.0267}{{\ttfamily arXiv:1305.0267
  [hep-ph]}}.
[Erratum: Nucl. Phys.B888,284(2014)].

\bibitem{Giudice:2000ex}
G.~F. Giudice, E.~W. Kolb, and A.~Riotto, ``{Largest temperature of the
  radiation era and its cosmological implications},''
  \href{http://dx.doi.org/10.1103/PhysRevD.64.023508}{{\em Phys. Rev.}
  {\bfseries D64} (2001) 023508},
\href{http://arxiv.org/abs/hep-ph/0005123}{{\ttfamily arXiv:hep-ph/0005123
  [hep-ph]}}.

\bibitem{Moroi:1994rs}
T.~Moroi, M.~Yamaguchi, and T.~Yanagida, ``{On the solution to the Polonyi
  problem with 0 (10-TeV) gravitino mass in supergravity},''
  \href{http://dx.doi.org/10.1016/0370-2693(94)01337-C}{{\em Phys. Lett.}
  {\bfseries B342} (1995) 105--110},
\href{http://arxiv.org/abs/hep-ph/9409367}{{\ttfamily arXiv:hep-ph/9409367
  [hep-ph]}}.

\bibitem{Kawasaki:1995cy}
M.~Kawasaki, T.~Moroi, and T.~Yanagida, ``{Constraint on the reheating
  temperature from the decay of the Polonyi field},''
  \href{http://dx.doi.org/10.1016/0370-2693(95)01546-9}{{\em Phys. Lett.}
  {\bfseries B370} (1996) 52--58},
\href{http://arxiv.org/abs/hep-ph/9509399}{{\ttfamily arXiv:hep-ph/9509399
  [hep-ph]}}.

\bibitem{Moroi:1999zb}
T.~Moroi and L.~Randall, ``{Wino cold dark matter from anomaly mediated SUSY
  breaking},'' \href{http://dx.doi.org/10.1016/S0550-3213(99)00748-8}{{\em
  Nucl. Phys.} {\bfseries B570} (2000) 455--472},
\href{http://arxiv.org/abs/hep-ph/9906527}{{\ttfamily arXiv:hep-ph/9906527
  [hep-ph]}}.

\bibitem{Gelmini:2006pw}
G.~B. Gelmini and P.~Gondolo, ``{Neutralino with the right cold dark matter
  abundance in (almost) any supersymmetric model},''
  \href{http://dx.doi.org/10.1103/PhysRevD.74.023510}{{\em Phys. Rev.}
  {\bfseries D74} (2006) 023510},
\href{http://arxiv.org/abs/hep-ph/0602230}{{\ttfamily arXiv:hep-ph/0602230
  [hep-ph]}}.

\bibitem{Garcia:2017tuj}
M.~A.~G. Garcia, Y.~Mambrini, K.~A. Olive, and M.~Peloso, ``{Enhancement of the
  Dark Matter Abundance Before Reheating: Applications to Gravitino Dark
  Matter},'' \href{http://dx.doi.org/10.1103/PhysRevD.96.103510}{{\em Phys.
  Rev.} {\bfseries D96} no.~10, (2017) 103510},
\href{http://arxiv.org/abs/1709.01549}{{\ttfamily arXiv:1709.01549 [hep-ph]}}.

\bibitem{Dudas:2017kfz}
E.~Dudas, T.~Gherghetta, Y.~Mambrini, and K.~A. Olive, ``{Inflation and
  High-Scale Supersymmetry with an EeV Gravitino},''
  \href{http://dx.doi.org/10.1103/PhysRevD.96.115032}{{\em Phys. Rev.}
  {\bfseries D96} no.~11, (2017) 115032},
\href{http://arxiv.org/abs/1710.07341}{{\ttfamily arXiv:1710.07341 [hep-ph]}}.

\bibitem{Drees:2017iod}
M.~Drees and F.~Hajkarim, ``{Dark Matter Production in an Early Matter
  Dominated Era},'' \href{http://dx.doi.org/10.1088/1475-7516/2018/02/057}{{\em
  JCAP} {\bfseries 1802} no.~02, (2018) 057},
\href{http://arxiv.org/abs/1711.05007}{{\ttfamily arXiv:1711.05007 [hep-ph]}}.

\bibitem{Garcia:2018wtq}
M.~A.~G. Garcia and M.~A. Amin, ``{Prethermalization production of dark
  matter},'' \href{http://dx.doi.org/10.1103/PhysRevD.98.103504}{{\em Phys.
  Rev.} {\bfseries D98} no.~10, (2018) 103504},
\href{http://arxiv.org/abs/1806.01865}{{\ttfamily arXiv:1806.01865 [hep-ph]}}.

\bibitem{Allahverdi:2018aux}
R.~Allahverdi and J.~K. Osi{\'n}ski, ``{Non-thermal Dark Matter from Modified
  Early Matter Domination},''
\href{http://arxiv.org/abs/1812.10522}{{\ttfamily arXiv:1812.10522 [hep-ph]}}.

\bibitem{Kaneta:2019zgw}
K.~Kaneta, Y.~Mambrini, and K.~A. Olive, ``{Radiative Production of Non-thermal
  Dark Matter},''
\href{http://arxiv.org/abs/1901.04449}{{\ttfamily arXiv:1901.04449 [hep-ph]}}.

\bibitem{Landau:1953um}
L.~D. Landau and I.~Pomeranchuk, ``{Limits of applicability of the theory of
  bremsstrahlung electrons and pair production at high-energies},''
{\em Dokl. Akad. Nauk Ser. Fiz.} {\bfseries 92} (1953) 535--536.

\bibitem{Migdal:1956tc}
A.~B. Migdal, ``{Bremsstrahlung and pair production in condensed media at
  high-energies},''
\href{http://dx.doi.org/10.1103/PhysRev.103.1811}{{\em Phys. Rev.} {\bfseries
  103} (1956) 1811--1820}.

\bibitem{Gyulassy:1993hr}
M.~Gyulassy and X.-n. Wang, ``{Multiple collisions and induced gluon
  Bremsstrahlung in QCD},''
  \href{http://dx.doi.org/10.1016/0550-3213(94)90079-5}{{\em Nucl. Phys.}
  {\bfseries B420} (1994) 583--614},
\href{http://arxiv.org/abs/nucl-th/9306003}{{\ttfamily arXiv:nucl-th/9306003
  [nucl-th]}}.

\bibitem{Arnold:2001ba}
P.~B. Arnold, G.~D. Moore, and L.~G. Yaffe, ``{Photon emission from
  ultrarelativistic plasmas},''
  \href{http://dx.doi.org/10.1088/1126-6708/2001/11/057}{{\em JHEP} {\bfseries
  11} (2001) 057},
\href{http://arxiv.org/abs/hep-ph/0109064}{{\ttfamily arXiv:hep-ph/0109064
  [hep-ph]}}.

\bibitem{Arnold:2001ms}
P.~B. Arnold, G.~D. Moore, and L.~G. Yaffe, ``{Photon emission from quark gluon
  plasma: Complete leading order results},''
  \href{http://dx.doi.org/10.1088/1126-6708/2001/12/009}{{\em JHEP} {\bfseries
  12} (2001) 009},
\href{http://arxiv.org/abs/hep-ph/0111107}{{\ttfamily arXiv:hep-ph/0111107
  [hep-ph]}}.

\bibitem{Arnold:2002ja}
P.~B. Arnold, G.~D. Moore, and L.~G. Yaffe, ``{Photon and gluon emission in
  relativistic plasmas},''
  \href{http://dx.doi.org/10.1088/1126-6708/2002/06/030}{{\em JHEP} {\bfseries
  06} (2002) 030},
\href{http://arxiv.org/abs/hep-ph/0204343}{{\ttfamily arXiv:hep-ph/0204343
  [hep-ph]}}.

\bibitem{Besak:2010fb}
D.~Besak and D.~Bodeker, ``{Hard Thermal Loops for Soft or Collinear External
  Momenta},'' \href{http://dx.doi.org/10.1007/JHEP05(2010)007}{{\em JHEP}
  {\bfseries 05} (2010) 007},
\href{http://arxiv.org/abs/1002.0022}{{\ttfamily arXiv:1002.0022 [hep-ph]}}.

\bibitem{Kurata:2012nf}
Y.~Kurata and N.~Maekawa, ``{Averaged Number of the Lightest Supersymmetric
  Particles in Decay of Superheavy Particle with Long Lifetime},''
  \href{http://dx.doi.org/10.1143/PTP.127.657}{{\em Prog. Theor. Phys.}
  {\bfseries 127} (2012) 657--664},
\href{http://arxiv.org/abs/1201.3696}{{\ttfamily arXiv:1201.3696 [hep-ph]}}.

\bibitem{Harigaya:2016vda}
K.~Harigaya, T.~Lin, and H.~K. Lou, ``{GUTzilla Dark Matter},''
  \href{http://dx.doi.org/10.1007/JHEP09(2016)014}{{\em JHEP} {\bfseries 09}
  (2016) 014},
\href{http://arxiv.org/abs/1606.00923}{{\ttfamily arXiv:1606.00923 [hep-ph]}}.

\bibitem{Mambrini:2013iaa}
Y.~Mambrini, K.~A. Olive, J.~Quevillon, and B.~Zaldivar, ``{Gauge Coupling
  Unification and Nonequilibrium Thermal Dark Matter},''
  \href{http://dx.doi.org/10.1103/PhysRevLett.110.241306}{{\em Phys. Rev.
  Lett.} {\bfseries 110} no.~24, (2013) 241306},
\href{http://arxiv.org/abs/1302.4438}{{\ttfamily arXiv:1302.4438 [hep-ph]}}.

\bibitem{Mambrini:2015vna}
Y.~Mambrini, N.~Nagata, K.~A. Olive, J.~Quevillon, and J.~Zheng, ``{Dark matter
  and gauge coupling unification in nonsupersymmetric SO(10) grand unified
  models},'' \href{http://dx.doi.org/10.1103/PhysRevD.91.095010}{{\em Phys.
  Rev.} {\bfseries D91} no.~9, (2015) 095010},
\href{http://arxiv.org/abs/1502.06929}{{\ttfamily arXiv:1502.06929 [hep-ph]}}.

\bibitem{Soni:2016gzf}
A.~Soni and Y.~Zhang, ``{Hidden SU(N) Glueball Dark Matter},''
  \href{http://dx.doi.org/10.1103/PhysRevD.93.115025}{{\em Phys. Rev.}
  {\bfseries D93} no.~11, (2016) 115025},
\href{http://arxiv.org/abs/1602.00714}{{\ttfamily arXiv:1602.00714 [hep-ph]}}.

\bibitem{Benakli:2017whb}
K.~Benakli, Y.~Chen, E.~Dudas, and Y.~Mambrini, ``{Minimal model of gravitino
  dark matter},'' \href{http://dx.doi.org/10.1103/PhysRevD.95.095002}{{\em
  Phys. Rev.} {\bfseries D95} no.~9, (2017) 095002},
\href{http://arxiv.org/abs/1701.06574}{{\ttfamily arXiv:1701.06574 [hep-ph]}}.

\bibitem{Dudas:2017rpa}
E.~Dudas, Y.~Mambrini, and K.~Olive, ``{Case for an EeV Gravitino},''
  \href{http://dx.doi.org/10.1103/PhysRevLett.119.051801}{{\em Phys. Rev.
  Lett.} {\bfseries 119} no.~5, (2017) 051801},
\href{http://arxiv.org/abs/1704.03008}{{\ttfamily arXiv:1704.03008 [hep-ph]}}.

\bibitem{Kang:2019izi}
Z.~Kang, ``{Slightly Ultra-violet Freeze-in a Hidden Gluonic Sector},''
\href{http://arxiv.org/abs/1901.10934}{{\ttfamily arXiv:1901.10934 [hep-ph]}}.

\end{thebibliography}\endgroup

\end{document}